# One software tool for testing square s-boxes

Dr. Eng. Nikolai Stoianov
Member of INDECT WP8 team of
Technical University of Sofia (TUS)
Technical University of Sofia
Sofia, Bulgaria
e-mail: nkl_stnv@tu-sofia.bg

*Abstract.* **An encryption technique is widely used to keep data confidential. Most of the block symmetric algorithms use substitution functions. Often this functions use so called S-BOX matrix. In this paper author presents one software tool for testing and measuring square s-boxes. Based of information theory functions for testing static and dynamic criteria are presented. These criterions are mathematically defined for square s-boxes. Two new criteria "private criteria" a proposed and pseudo codes for they creation and testing are presented.**

## I. INTRODUCTION

The creation, processing, delivering and analysis of information always have been a key element of any strategy and operation. With the progress of communication and information technologies, during the last decade of XX and from the beginning of XXI centuries the role of these systems and technologies has increased significantly. One of main used technology for data confidentiality is encryption. The development of the information technology and in particular of the multiple increase of the processors' speed led to a necessity to revise the encryption algorithms in use. [10],[11],[16].

## II. S-BOX REQUIREMENTS

The defined requirements, to which every S-BOX shall comply are dictated by the need the algorithm to be reliable both to differential and linear crypto analysis and to have stability of the algebraic transformation like interpolating attacks e.g. Considering those conditions the following requirements to the S-BOXES are defined [1],[2],[3],[4],[5],[6],[7],[8],[9],[11],[12],[13]:

- General criteria:
    o Invertability;
    o Minimization of the correlation between the linear combination of the input bytes and the linear combination of the output bytes;
    o Complexity in the algebraic presentation in the Galois Field $GF(2^8)$;
    o Non-linearity;
    o Stability and reliability to differential crypto analysis;
    o Avalanche effect – Probability, direct and minimal.
- Additional (Side) criteria:
    o Independence between the input and output data;
    o Independence between the output and input data;
    o Independence between the output and output data;
- Dynamic criteria:
    o Dynamic Independence between the input and output data;
    o Dynamic Independence between the output and input data;
    o Dynamic Independence between the output and output data;
- Private criteria:
    o Completeness – the generated and the input S-BOX will be complete (there is value for every cell of the table);
    o Non-contradiction – The S-BOX which is tested will have only one non repeated value in each cell of the table.

Author's team develops a tool via which the statistical, dynamic and private criteria for each generated and tested S-BOX will be checked as well as inverse S-BOXES will be created and processing of input data (plain text or seed) with a specific S-BOX and record in a file.

## III. FUNCTIONALITIES OF THE SOFTWARE TOOL

The Software product „S-BOX Tests" is functionally divided into two main fields:

- Data processing;
- Checks and Tests.

### A. Data Processing

This functionality of the software tool allows us to conduct the following activities:

1. **To receive an inverse (reversed) S-BOX and save the result in a file.**

As an input data a text file with a prototype of a S_BOX with a size n x n is used, where the values are delimited by comma. Based of the created file an inverse S-BOX is created and this new S-BOX is saved into another text file, where the values are also separated by commas.

Pseudo code of the function "Inv S-BOX" is the following:

```
Begin
  Open_Input_S-BOX_File;
  Read_Data_From_Input_S-BOX_File;
  Empty(Array_InvS-BOX[0..n,0..n]); //fill with null
  Fill(Array_S-BOX[0..n,0..n]);
  For i=0 to n do
    For j=0 to n do
      Begin
       XY=Array_S-BOX[j,y]
       Array_InvS-BOX[XY]=ij
      End;
  Show_Result_On_Screen;
  Save_Output_InvS-BOX_File;
End.
```

**2. Input information flow (file) processing via selected S-BOX and record of the result into a file;**

As input data selected S-BOX in the format of a text file with values separated by comma is used. The input flow (plain text or seed) is in a file of Byte type. After the data processing the result is saved in input file of Byte Type.

The pseudo code of the function "F_to_S-BOX" is the following:

```
Begin
  Open_Input_S-BOX_File;
  Open_Input_Text_File;
  Fill_Array_S-BOX[0..n,0..n];
  Read_Byte_From_Text_File(A);
  B=Make_Subbstitution(A,Array_S-BOX);
  Write_Byte_To_Output_SubText_File(B);
  Save_Output_SubText_File;
End.
```

**B. Checks and Tests**

*Checks*

The functions developed in the "Checks" module shall identify if the data provided for the S-BOX is correct. Having in mind that the statistic and dynamic tests take time, it is reasonable first to check each generated (delivered) S-BOX for correctness.

**1. For completeness.**

This function is incorporated in the software in order to check if the check subject S-BOX is complete i.e. if all the cells of the matrix have values. As an input data a text file with values delimited by commas. The pseudo code of the "Completeness" function is:

```
Begin
  Open_Input_S-BOX_File;
  Read_Data_From_Input_S-BOX_File;
  Fill(Array_S-BOX[0..n,0..n]);
  For i=0 to n do
    For j=0 to n do
        If Array_S-BOX[0..n,0..n]=null then
Result=False
        Else Result=True;
End.
```

**2. For non-contradiction**

Via the function "Non-contradiction" every S-BOX is checked against duplication of values in the cells of the matrix. As input data a text file with values delimited by commas is used. The result of this function is Boolean (False or True). The pseudo code of this function is:

```
Begin
  Open_Input_S-BOX_File;
  Read_Data_From_Input_S-BOX_File;
  Fill(Array_S-BOX[0..n,0..n]);
  For i=0 to n do
    For j=0 to n do
      Begin
        A= Array_S-BOX[i,j];
        Check (A, Array_S-BOX])
      End;
End.
```

Based on the Theory of Information and as per [3],[12],[13],[14],[15] the goal of each good or ideal S-BOX is to minimize the information, which might be received about unknown input data transformed via S-BOX on the base of other known input and output data, i.e. if accidental variable X describes some known data and the accidental variable Y respectively the unknown data, so:

$$I(X;Y) = H(X) + H(Y) - H(X,Y)$$

$$I(X;Y) = H(X) - H(Y \mid X) = 0 \text{ for an ideal S-BOX,}$$

where: $H(X) = -\sum_{i=1}^{n} P_i . \log(P_i)$,

$H(X \mid Y) = H(X,Y) - H(Y)$,

$H(X,Y) = -\sum\sum P_{ij}.\log(P_{ij}) = \sum\sum P_{ij}.\log(P(y_j \mid x_i))$

When we use binary system

$$H(X) = \sum_{i=1}^{n} P(x_i).\log_2\left(\frac{1}{P(x_i)}\right)$$

Each S-BOX is with the size *n x n*, i.e. it is square.

*Static tests*

**1. Independence between input and output data;**

For the purposes of the software tests we assume the following:

Each S-BOX meets the requirement for independence between input and output data of the row *r*, where *r<n* if:

$$P(y_i \mid a_1x_1, a_2x_2,...,a_nx_n) = P(y_i)$$

for
$$\forall x_i, y_j, a_k \mid 1 \leq i,j,k \leq n; (x_i, y_j, a_k) \in \{0,1\};$$
$$A = [a_1, a_2, ..., a_n]$$

The number of known input data is *r*, i.e. *Count(A=1)=r* and $a_k=0$ which means that $x_k$ is unknown, and with $a_k=1- x_k$ is known.

For testing the independency between the input and output data the function "Stat I/O" is developed. As a function input the following items are used:

- The tested S-BOX as a text file with data delimited by comma;
- A text prepared in advance (plain text or seed);
- The resultant text after its processing (substitution) with the tested S-BOX;
- The number of the known input bytes (*r*).

## 2. Independence between output and input data;

For the purpose of the software tests we assume the following:

Each S-BOX meets the requirement for independence between output and input data from the row *r*, where *r<n* if:

$$P(x_i | a_1 y_1, a_2 y_2, ..., a_n y_n) = P(x_i)$$

for
$$\forall x_i, y_j, a_k \mid 1 \leq i,j,k \leq n; (x_i, y_j, a_k) \in \{0,1\};$$
$$A = [a_1, a_2, ..., a_n]$$

the number of the known input data is *r*, i.e. *Count(A=1)=r* and $a_k=0$ which means that $y_k$ is unknown and with $a_k=1- y_k$ is known.

For testing the independence between the input and output data the function "Stat O/I" is developed. For the input of this function the following items are used:

- The tested S-BOX as a text file with data delimited by comma;
- A text prepared in advance (plain text or seed);
- A resultant text after its processing (substitution) with the tested S-BOX;
- The number of the known output bytes (*r*).

## 3. Independence between the output and output data;

For the purpose of the software tests we assume the following:

Each S-BOX meets the requirement for independence between the output and output data of the row *r*, where *r<n* if:

$$P(y_i | a_1 y_1, a_2 y_2, ..., a_n y_n) = P(y_i)$$

for
$$\forall y_j, a_k \mid 1 \leq j,k \leq n; (y_j, a_k) \in \{0,1\};$$
$$A = [a_1, a_2, ..., a_n]$$

The number of the known input data is *r*, i.e. *Count(A=1)=r* and with $a_k=0$ which means that $x_k$ is unknown and with $a_k=1- x_k$ is known.

Also

$$H(Y) = \begin{cases} H(X), & \text{if } H(X) \leq n \\ n, & \text{if } H(X) > n \end{cases}$$

where $X=[x_1, x_2, ...x_n]$ and $Y=[y_1, y_2, ..., y_n]$

For testing the independence between the output and output data the function "Stat O/O" is developed. As an input for this function the following items are used:

- The tested S-BOX as a text file with data delimited by comma;
- A text prepared in advance (plain text or seed);
- The resultant text after its processing (substitution) with the tested S-BOX;
- The number of the known bytes (*r*).

*Dynamic tests*

## 1. The Dynamic independence between the input and output data;

For the purpose of the software tests we assume the following:

Each S-BOX meets the requirements for independence between the input and output data of the row *r*, with *r<n* if:

$$P(\Delta y_i | a_1 \Delta x_1, a_2 \Delta x_2, ..., a_n \Delta x_n) = P(\Delta y_i)$$

for
$$\forall \Delta x_i, \Delta y_j, a_k \mid 1 \leq i,j,k \leq n; (\Delta x_i, \Delta y_j, a_k) \in \{0,1\};$$
$$A = [a_1, a_2, ..., a_n]$$

the number of known input data is *r*, i.e. *Count(A=1)=r* and $a_k=0$, which means that $\Delta x_k$ is unknown, and with $a_k=1- \Delta x_k$ is known.

For testing the independence between the input and output data the function "Dynamic I/O" is developed. For the function's input the following items are used:

- The tested S-BOX as a text file with data delimited by comma;
- A text prepared in advance (plain text or seed);
- A second text prepared in advance 2 (plain text or seed);
- A resultant text 1 after its processing (substitution) with the tested S-BOX;

- A resultant text 2 after its processing (substitution) with the tested S-BOX;
- The number of the known input bytes ($r$).

2. **Dynamic independence between the output and input data;**

For the purpose of the software tests we assume the following:

Each S-BOX meets the requirement for independence between the output and input data of the row $r$, with $r<n$ if:

$$P(\Delta x_i \mid a_1\Delta y_1, a_2\Delta y_2, ..., a_n\Delta y_n) = P(\Delta x_i)$$

for $\forall \Delta x_i, \Delta y_j, a_k \mid 1 \leq i,j,k \leq n; (\Delta x_i, \Delta y_j, a_k) \in \{0,1\};$
$A = [a_1, a_2, ..., a_n]$

the number of the known input data is $r$, i.e. $Count(A=1)=r$ with $a_k=0$ which means that $\Delta y_k$ is unknown and with $a_k=1$ - $\Delta y_k$ is known.

For testing the independence between the input and output data the function "Dynamic O/I" is developed. As an input for this function the following items ate used:

- The tested S-BOX as a text file with data delimited by comma;
- A text 1 prepared in advance (plain text or seed);
- A second text 2 prepared in advance (plain text or seed);
- A resultant text 1 after its processing (substitution) with the tested S-BOX;
- A resultant text 2 after its processing (substitution) with the tested S-BOX;
- The number of the known input bytes ($r$).

3. **Dynamic independence between the output and output data;**

For the purpose of the software tests we assume the following:

Each S-BOX meets the requirement for independence between the output and output data of the row $r$, with $r<n$ if:

$$P(\Delta y_i \mid a_1\Delta y_1, a_2\Delta y_2, ..., a_n\Delta y_n, \Delta x_1, \Delta x_2, ..., \Delta x_n) = P(\Delta y_i \mid \Delta x_1, \Delta x_2, ..., \Delta x_n)$$

for
$\forall \Delta x_i, \Delta y_j, a_k \mid 1 \leq i,j,k \leq n; (\Delta x_i, \Delta y_j, a_k) \in \{0,1\}$
$A = [a_1, a_2, ..., a_n]$

the number of the known input data is $r$, i.e. $Count(A=1)=r$ and with $a_k=0$ which means that $\Delta x_k$ is unknown and with $a_k=1$ - $\Delta x_k$ is known.

And also

$$H(\Delta Y) = \begin{cases} H(\Delta X), & \text{if } H(\Delta X) \leq n \\ n, & \text{if } H(\Delta X) > n \end{cases}$$

where $\Delta X=[\Delta x_1, \Delta x_2, ..., \Delta x_n]$ and $Y=[\Delta y_1, \Delta y_2, ..., \Delta y_n]$

For testing the independence between the input and input data the function "Dynamic O/O" is developed. As input data for this function the following items are used:

- The tested S-BOX as a text file with data delimited by comma;
- Preliminary prepared text 1 (plain text or seed);
- Preliminary prepared text 2 (plain text or seed);
- Resultant text 1 after its processing (substitution) with the tested S-BOX;
- Resultant text 2 after its processing (substitution) with the tested S-BOX;
- The number of the known input bytes ($r$).

**Remark:** The results received from the processing of S-BOXES given in the algorithm AES is accepted as basic and all the received results (for correctness check and functional tests) are compared with those from the S-BOX$_{AES}$.

**Acknowledgment** The work presented in this paper has been performed in the framework of the EU Project INDECT (*Intelligent information system supporting observation, searching and detection for security of citizens in urban environment*) — grant agreement number: 218086.